\documentclass[12pt,leqno]{amsart}
\usepackage{amssymb}
\usepackage{amsmath,amssymb}
\newtheorem{proposition}{Proposition}[section]

\begin{document}
\theoremstyle{plain}
\newtheorem{MainThm}{Theorem}
\newtheorem{thm}{Theorem}[section]
\newtheorem{clry}[thm]{Corollary}
\newtheorem{prop}[thm]{Proposition}
\newtheorem{lem}[thm]{Lemma}
\newtheorem{deft}[thm]{Definition}
\newtheorem{hyp}{Assumption}
\newtheorem*{ThmLeU}{Theorem (J.~Lee, G.~Uhlmann)}

\theoremstyle{definition}
\newtheorem{rem}[thm]{Remark}
\newtheorem*{acknow}{Acknowledgments}
\numberwithin{equation}{section}
\newcommand{\eps}{{\varphi}repsilon}
\renewcommand{\d}{\partial}
\newcommand{\re}{\mathop{\rm Re} }
\newcommand{\im}{\mathop{\rm Im}}
\newcommand{\R}{\mathbf{R}}
\newcommand{\C}{\mathbf{C}}
\newcommand{\N}{\mathbf{N}}
\newcommand{\D}{C^{\infty}_0}
\renewcommand{\O}{\mathcal{O}}
\newcommand{\dbar}{\overline{\d}}
\newcommand{\supp}{\mathop{\rm supp}}
\newcommand{\abs}[1]{\lvert #1 \rvert}
\newcommand{\csubset}{\Subset}
\newcommand{\detg}{\lvert g \rvert}
\newcommand{\ppp}{\partial}
\newcommand{\dd}{\mbox{div}\thinspace}

\title
[Calder\'on problem for Maxwell's equations]
{Calder\'on problem for Maxwell's equations in two dimensions}

\author{
O.~Yu.~Imanuvilov and \,
M.~Yamamoto }
\thanks{ Department of Mathematics, Colorado State
University, 101 Weber Building, Fort Collins, CO 80523-1874, U.S.A.
e-mail: {\tt oleg@math.colostate.edu}.
Partially supported by NSF grant DMS 1312900}\,
\thanks{ Department of Mathematical Sciences, The University
of Tokyo, Komaba, Meguro, Tokyo 153, Japan e-mail:
myama@ms.u-tokyo.ac.jp}

\date{}

\maketitle

\begin{abstract}
We prove the global uniqueness in determination of
the conductivity, the permeability and the permittivity of
two dimensional Maxwell's equations by partial Dirichlet-to-Neumann
map limited to an arbitrary subboundary.
\end{abstract}

Let $\Omega \subset \R^3$ be a bounded domain with smooth boundary $\d\Omega$,
and
${\vec\nu} = (\nu_1,\nu_2,\nu_3) \in \R^3$ be a unit outward normal vector
to $\partial\Omega$ and let $i = \sqrt{-1}$.
Let $E=(E_1,E_2,E_3)$ be the electric field, $H=(H_1,H_2,H_3)$
the magnetic field, $\sigma$ be the conductivity, $\mu$ the permeability
and $\epsilon$ permittivity.
In this paper we assume that $E(x), H(x), \sigma(x), \mu(x), \epsilon(x)$ are
independent of the third component $x_3$ of $x = (x_1,x_2,x_3) \in \R^3$.
Then we embed $\Omega$ in $\R^2$ and regard
$\Omega$ as a domain in $\R^2$.  Then ${\vec\nu} = (\nu_1,\nu_2,0)$
and
$$
{\vec \nu\times E}
= \left ( \begin{matrix} \nu_2E_3\\
-\nu_1E_3\\
\nu_1E_2 - \nu_2E_1
\end{matrix}\right)  \quad \mbox{on $\partial\Omega$}
$$
and Maxwell's equations are given by
\begin{equation}\label{3}
L_{1,\mu,\gamma}(x,D)(E,H):=\left ( \begin{matrix} \partial_{x_2}E_3\\
-\partial_{x_1}E_3\\\partial_{x_1}E_2-\partial_{x_2}E_1   \end{matrix}
\right)-i\omega\mu\left (\begin{matrix}H_1\\H_2\\
H_3\end{matrix} \right)=0,\quad\mbox{in}\,\,\Omega,
\end{equation}
and
\begin{equation}\label{4}
L_{2,\mu,\gamma}(x,D)(E,H):=\left ( \begin{matrix} \partial_{x_2}H_3\\
-\partial_{x_1}H_3\\\partial_{x_1}H_2-\partial_{x_2}H_1   \end{matrix}
\right)+i\omega\gamma\left (\begin{matrix}E_1\\E_2\\
E_3\end{matrix} \right)=0,\quad\mbox{in}\,\,\Omega.
\end{equation}
Here and henceforth we set
$$
\gamma=\epsilon+\frac{i\sigma}{\omega}
$$
and
$$
L_{\mu,\gamma}(x,D)(E,H) = (L_{1,\mu,\gamma}(x,D)(E,H),
L_{2,\mu,\gamma}(x,D)(E,H)).
$$

Let $\widetilde \Gamma$ be some fixed open subset of $\partial\Omega$ and
$\Gamma_0=\partial\Omega\setminus\widetilde \Gamma.$
Consider the following Dirichlet-to-Neumann map
\begin{equation}\label{2}
\Lambda_{\mu,\gamma}f = {\vec\nu}\times H
= \left (\begin{matrix} \nu_2H_3\\-\nu_1H_3\\
\nu_1H_2-\nu_2H_1
\end{matrix}\right)\quad\mbox{on}\,\,\widetilde \Gamma,
\end{equation}
where
$$
L_{\mu,\gamma}(x,D) (E,H)=0 \quad\mbox{in}\,\,\Omega,\quad
{\vec \nu}\times E\vert_{\Gamma_0}=0, \quad {\vec \nu}
\times E\vert_{\widetilde \Gamma}=f.
$$

In general for some  values of the parameter $\omega$, the boundary value problem
\begin{equation}\label{general}
L_{\mu,\gamma}(x,D) (E,H)=0\quad\mbox{in}\,\,\Omega,\quad \vec
\nu\times E\vert_{\Gamma_0}=0, \quad \vec \nu\times E\vert
_{\widetilde{\Gamma}}=f
\end{equation}
may not have a solution for some $f$.
By $D_{\mu,\gamma}$ we denote the set of functions
$f\in H^1(\widetilde\Gamma)$ such that there exists at least one solution
to (\ref{general}).  As for the mathematical theory on the boundary
value problem for Maxwell's equations, we refer for example to Dautray and
Lions \cite{DL}.

In general for some $f\in D_{\mu,\gamma}$, there exists more than one
solutions.  In that case as the value of $\Lambda_{\mu,\gamma}f$,
we consider the set of all  functions $\vec\nu\times H$ where the
pairs $(E,H)$ are the all possible solutions to (\ref{general}).
Thus our definition of the Dirichlet-to-Neumann map is different form
the classical one, and we have to specify the conception of the equality of
the Dirichlet-to-Neumann maps.

{\bf Definition.}
{\it We say that the Dirichelt-to-Neumann maps $\Lambda_{\mu_1,\gamma_1}$
and $\Lambda_{\mu_2,\gamma_2}$ are equal
if $D_{\mu_1,\gamma_1}\subset D_{\mu_2,\gamma_2}$
and for any pair $(E,H)$ which solves
\begin{equation}\nonumber
L_{\mu_1,\gamma_1}(x,D) (E,H)=0\quad\mbox{in}\,\,\Omega,\quad \vec \nu
\times E\vert_{\Gamma_0}=0, \quad \vec \nu\times E\vert_{\widetilde\Gamma}
= f,
\end{equation}
there exists a pair $(\widetilde E,\widetilde H)$ which solves
\begin{equation}\nonumber
L_{\mu_1,\gamma_1}(x,D) (\widetilde E,\widetilde H)=0
\quad\mbox{in}\,\,\Omega,\quad \vec \nu\times \tilde E\vert_{\Gamma_0}=0,
\quad \vec \nu\times\tilde  E\vert_{\widetilde\Gamma}=f
\end{equation}
and
$$
\vec\nu\times H=\vec\nu\times \tilde H\quad \mbox{on}\quad
{\widetilde\Gamma}.
$$
}

Then we can state our main result:
\\
\vspace{0.2cm}
\\
{\bf Theorem}
{\it Let $\mu_j,\epsilon_j,\sigma_j\in C^5(\bar \Omega)$ for $j\in\{1,2\}$
and $\mu_j, \epsilon_j$ be the positive functions on $\bar \Omega.$
Suppose that $\Lambda_{\mu_1,\gamma_1} = \Lambda_{\mu_2,\gamma_2}$ and
$$
\mu_1-\mu_2=\frac{\partial\mu_1}{\partial\nu}-\frac{\partial\mu_2}{\partial\nu}=\gamma_1-\gamma_2=
\frac{\partial\gamma_1}{\partial\nu}-\frac{\partial\gamma_2}{\partial\nu}=0\quad \mbox{on}\,\,\tilde \Gamma\,\, \mbox{and} \quad\frac{\partial (\root\of{\gamma_1}-\root\of{\gamma_2})}{\partial\nu}=0 \quad \mbox{on}\,\,\Gamma_0.
$$
Then $\mu_1=\mu_2$  and $\gamma_1=\gamma_2.$
}

See Caro, Ola and Salo \cite{POM} and Ola, P\" aiv\"arinta and Somersalo
\cite{OPS} for the uniquenesss results for Maxwell's equations in three
dimensions.  In the two dimensional case, we can reduce Maxwell's equations
to the conductivity equation with zeroth order term and apply the
uniqueness result in Imanuvilov, Uhlmann and Yamamoto \cite{IUY1} to
prove the theorem.

{\bf Proof.} First we observe that the system (\ref{3}), (\ref{4}) can be separated into the two independent systems of partial differential equations. The first system has the form
\begin{equation}\label{8}
H_1=\frac{1}{i\omega\mu}\partial_{x_2}E_3,\quad H_2=-\frac{1}{i\omega\mu}\partial_{x_1}E_3,\quad \partial_{x_1}H_2-\partial_{x_2}H_1=-i\omega\gamma E_3\quad\mbox{in}\,\,\Omega.
\end{equation}
The second system can be written as

\begin{equation}\label{9}
E_1=-\frac{1}{i\omega\gamma}\partial_{x_2}H_3,\quad E_2=\frac{1}{i\omega\gamma}\partial_{x_1}H_3,\quad \partial_{x_1}E_2-\partial_{x_2}E_1=i\omega\mu H_3\quad\mbox{in}\,\,\Omega.
\end{equation}

After we plug  into the third equation of (\ref{8}) expressions for $H_1$ and $H_2$ from the first two equations we obtain
\begin{equation}\label{10}
L_{1,\mu,\gamma}(x,D)E_3=\mbox{div}\,\left (\frac{1}{i\omega\mu}\nabla E_3\right  )-i\omega\gamma E_3=0\quad\mbox{in}\,\,\Omega.
\end{equation}

Similarly, from (\ref{9}) we obtain

\begin{equation}\label{11}
L_{2,\gamma,\mu}(x,D)H_3=-\mbox{div}\,\left (\frac{1}{i\omega \gamma}\nabla H_3\right)+i\omega\mu H_3=0\quad\mbox{in}\,\,\Omega.
\end{equation}

Observe that in order to solve the system (\ref{8}) it suffices to solve the conductivity equation (\ref{10}) and then determine the first to components of the magnetic field $H$ by formulae
\begin{equation}\label{PP}
H_1=\frac{1}{i\omega\mu}\partial_{x_2}E_3,\quad H_2=-\frac{1}{i\omega\mu}\partial_{x_1}E_3.
\end{equation}
Finally, using equation (\ref{PP}) in (\ref{10}) to eliminate $E_3$ we obtain the last equation in (\ref{8}).

Similarly in order to construct solution to (\ref{9}) we solve  the  conductivity equation (\ref{11}) for $H_3$ and then determine the first two components of the electric field $E$ by formulae
\begin{equation}
E_1=\frac{1}{i\omega \gamma}\partial_{x_2}H_3,\quad E_2=-\frac{1}{i\omega \gamma}\partial_{x_1}H_3 .
\end{equation}
Finally, using these formulae  in (\ref{11}) to eliminate $H_3$ we obtain the last equation in (\ref{9}).

Next, we claim that if the Dirichlet-to-Neuman map is given we can recover the following

{\bf A}) The Dirichlet-to-Neumann map:
\begin{equation}\label{12}
\Lambda_{1,\mu,\gamma} f=\frac{\partial u}{\partial \nu}\vert_{\tilde \Gamma},
\end{equation}
where
\begin{equation}\label{12'}
\quad L_{1,\mu,\gamma}(x,D)u=0\quad\mbox{in}\,\,\Omega,\quad u\vert_{\Gamma_0}=0,\quad u\vert_{\tilde \Gamma}=f.
\end{equation}
Let $u$ be some solution to equation (\ref{12'}). We set $E_3=u, q_1=\nu_2E_3, q_2=-\nu_1E_3.$ Let $q_3=0$ and $H_3=E_1=E_2=0.$ Finally the functions $H_1$ and $H_2$ are given by (\ref{PP}).
Then
$$\vec \nu\times  E\vert_{\Gamma_0}=0
$$
and
$$
\left( \begin{matrix} \nu_2E_3\\-\nu_1E_3\\\nu_1E_2-\nu_2E_1 \end{matrix}\right)=\mbox{\bf q}=\left (\begin{matrix}q_1\\q_2\\q_3 \end{matrix}\right)\quad\mbox{on}\,\,\tilde \Gamma
$$
and the following is known:
$$
\Lambda_{\mu,\gamma} \mbox{\bf q}=\left (\begin{matrix} \nu_2H_3\\-\nu_1H_3\\\nu_1H_2-\nu_2H_1 \end{matrix}\right)\quad\mbox{on}\,\,\tilde \Gamma.
$$
The short computations imply
$$
\nu_1H_2-\nu_2H_1=\frac{1}{i\omega\mu}\frac{\partial E_3}{\partial\nu}=\frac{1}{i\omega\mu}\frac{\partial u}{\partial\nu}\quad\mbox{on}\,\,\tilde \Gamma.
$$
Since the  trace of the function $\mu$ is known on $\tilde \Gamma$ we can determine $\frac{\partial u}{\partial\nu}.$

{\bf B}) The Neumann-to-Dirichlet map
\begin{equation}\label{13}
\Lambda_{2,\gamma,\mu} f=u\vert_{\tilde\Gamma},
\end{equation}
where
\begin{equation}\label{13'}
\quad L_{2,\gamma,\mu}(x,D)u=0\quad\mbox{in}\,\,\Omega,\quad \frac{\partial u}{\partial\nu}\vert_{\Gamma_0}=0,\quad \frac{\partial u}{\partial\nu}\vert_{\tilde \Gamma}=f.
\end{equation}
Let $u$ be some solution to equation (\ref{13'}). We set $H_3=u$  and $ E_1=\frac{1}{i\omega \gamma}\partial_{x_2}H_3,\quad E_2=-\frac{1}{i\omega \gamma}\partial_{x_1}H_3 $ and $H_1=H_2=E_3\equiv 0.$ Let us make a some choice of the function $\mbox{\bf q}=(0,0, f)$. Then $\nu_1E_2-\nu_2E_1=-\nu_1\frac{1}{i\omega \gamma}\partial_{x_1}H_3-\nu_2\frac{1}{i\omega \gamma}\partial_{x_2}H_3=-\frac{1}{i\omega \gamma}\frac{\partial H_3}{\partial\nu}=-\frac{1}{i\omega \gamma}\frac{\partial u}{\partial\nu}.$
Since the traces of $\epsilon$ and $\sigma$ on $\tilde \Gamma$ are known the choice of $f$ uniquely determines $\nu_1E_2-\nu_2E_1$ on $\tilde \Gamma.$
Then from the Dirichlet-to-Neumann map (\ref{2}) we determine the traces of the functions $\nu_2 H_3$ and $\nu_1 H_3$ on $\tilde \Gamma.$
Hence $H_3=u$ is determined on $\tilde \Gamma.$

Consider the following Dirichlet-to-Neumann map

\begin{equation}\label{14}
\Lambda_{1,\gamma,\mu} f=\frac{\partial u}{\partial \nu}\vert_{\tilde \Gamma},
\end{equation}
where
\begin{equation}\label{14'}
\quad L_{2,\gamma,\mu}(x,D)u=0\quad\mbox{in}\,\,\Omega,\quad \frac{\partial u}{\partial\nu}\vert_{\Gamma_0}=0,\quad u\vert_{\tilde \Gamma}=f.
\end{equation}

We have
\begin{proposition} If the Neumann-to-Dirichlet map (\ref{13}), (\ref{13'}) is  know the Dirichlet-to-Neumann map (\ref{14}), (\ref{14'}) is also known.
\end{proposition}
{\bf Proof.} Indeed let for some function $\tilde g$ there exists a solution to the boundary value problem (\ref{14'}). We denote such a solution as $u_1$. Let $\frac{\partial u_1}{\partial\nu}\vert_{\tilde \Gamma}=f.$ Since $\Lambda_{2,\gamma_1,\mu_1}= \Lambda_{2,\gamma_2,\mu_2}$ there exists solution $u_2$ to problem (\ref{13'}) such that
$$
\frac{\partial u_1}{\partial\nu}\vert_{\tilde \Gamma}=\frac{\partial u_2}{\partial\nu}\vert_{\tilde \Gamma}  =f\quad \mbox{and}\quad u_1\vert_{\tilde \Gamma}= u_2\vert_{\tilde \Gamma}=g.
$$
The proof of the Proposition is complete.
$\blacksquare$

Making the change of unknown in (\ref{12'}) as $u=\root\of {\mu}w$ and in (\ref{14'}) as $u=\root\of{\gamma}w$ we have
$$
\tilde L_{1,\mu,\gamma}(x,D)w=\Delta w+(\omega^2\gamma\mu+\frac{\Delta \root\of\mu}{\root\of{\mu}})w=0\quad\mbox{in}\,\,\Omega
$$
and
$$
\tilde L_{2,\gamma,\mu}(x,D)w=\Delta w+(\omega^2\gamma\mu+\frac{\Delta \root\of\gamma}{\root\of{\gamma}})w=0\quad\mbox{in}\,\,\Omega .
$$
The above change of variables preserve the Dirichlet-to-Neumann map (\ref{12}), (\ref{12'}) and transform the Dirichlet-to-Neumann map (\ref{14}), (\ref{14'}) into the following one

\begin{equation}\label{141}
\Lambda_{3,\gamma,\mu} f=\frac{\partial u}{\partial \nu}\vert_{\tilde \Gamma},
\end{equation}
where
\begin{equation}\label{141'}
\quad \tilde L_{2,\gamma,\mu}(x,D)u=0\quad\mbox{in}\,\,\Omega,\quad (\frac{\partial u}{\partial\nu}+au)\vert_{\Gamma_0}=0,\quad u\vert_{\tilde \Gamma}=f,
\end{equation}
and $a=-\frac{\partial}{\partial\nu}\root\of{\gamma}.$

Since the potentials of the Sch\"odinger operator can be determined from  the partial Dirichlet-to-Neumann maps for the two Maxwell's systems with the same Dirichlet-to-Neumann maps we obtain
\begin{equation}\label{16}
(\omega^2(\epsilon_1+\frac{i\sigma_1}{\omega}){\mu_1}+\frac{\Delta \root\of{\mu_1}}{\root\of{\mu_1}})=(\omega^2(\epsilon_2+\frac{i\sigma_2}{\omega}){\mu_2}+\frac{\Delta \root\of{\mu_2}}{\root\of{\mu_2}})
\end{equation}
and

\begin{equation}\label{17'}
\omega^2{(\epsilon_1+\frac{i\sigma_1}{\omega})}\mu_1+\frac{\Delta \root\of{(\epsilon_1+\frac{i\sigma_1}{\omega})}}{\root\of{(\epsilon_1+\frac{i\sigma_1}{\omega})}}=
\omega^2{(\epsilon_2+\frac{i\sigma_2}{\omega})}\mu_2+\frac{\Delta \root\of{(\epsilon_2+\frac{i\sigma_2}{\omega})}}{\root\of{(\epsilon_2+\frac{i\sigma_2}{\omega})}}.
\end{equation}

Denote
$r=(\root\of{\mu_1}-\root\of{\mu_2},\root\of{(\epsilon_1+\frac{i\sigma_1}{\omega})}-\root\of{(\epsilon_2+\frac{i\sigma_2}{\omega})}).$

By (\ref{16}), (\ref{17'}) there exist a matrices $A=(A_1,A_2,A_3),B\in L^\infty(\Omega)$ such that
$$
\Delta r+(A,\nabla r)+Br=0\quad\mbox{in}\,\,\Omega, \quad r\vert_{\tilde \Gamma}=\frac{\partial r}{\partial\nu}\vert_{\tilde \Gamma}=0.
$$

By the uniqueness theorem for the Cauchy problem for the such a system we have $r\equiv 0.$
The proof of the theorem is complete. $\blacksquare$

\end{document}